\documentclass[twocolumn]{aastex61}

\usepackage{graphicx}
\usepackage[varg]{txfonts}
\usepackage{natbib}
\usepackage{color}

\received{}
\revised{}
\accepted{}

\submitjournal{ApJ}

\shorttitle{ALMA polar brightening}
\shortauthors{Selhorst et al.}

\begin{document}

\title{SOLAR POLAR BRIGHTENING and RADIUS at 100 and 230 GHz observed by ALMA}

\correspondingauthor{Caius L. Selhorst}
\email{caiuslucius@gmail.com}

\author{Caius L. Selhorst}
\affiliation{NAT - N\'ucleo de Astrof\'isica Te\'orica, Universidade Cruzeiro do Sul, S\~ao Paulo, SP, Brazil}

\author{Paulo J. A. Sim\~oes}
\affiliation{SUPA School of Physics and Astronomy, University of Glasgow, G12 8QQ, UK}
\affiliation{CRAAM, Universidade Presbiteriana Mackenzie, S\~ao Paulo, SP 01302-907, Brazil}

\author{Roman Braj\v sa}
\affiliation{Hvar Observatory, Faculty of Geodesy, University of Zagreb, Zagreb, Croatia}

\author{Adriana Valio}
\affiliation{CRAAM, Universidade Presbiteriana Mackenzie, S\~ao Paulo, SP 01302-907, Brazil}

\author{C. G. Gim\'enez de Castro}
\affiliation{CRAAM, Universidade Presbiteriana Mackenzie, S\~ao Paulo, SP 01302-907, Brazil}
\affiliation{IAFE, Universidad de Buenos Aires/CONICET, Buenos Aires, Argentina}

\author{Joaquim E. R. Costa}
\affiliation{CEA, Instituto Nacional de Pesquisas Espaciais, S\~ao Jos\'e dos Campos, SP,  Brazil}

\author{Fabian Menezes}
\affiliation{CRAAM, Universidade Presbiteriana Mackenzie, S\~ao Paulo, SP 01302-907, Brazil}

\author{Jean Pierre Rozelot} 
\affiliation{Universit\'e  C\^ote d'Azur,  77 chemin des Basses Mouli\`eres, 06130 Grasse, France}

\author{Antonio S. Hales}
\affiliation{Joint ALMA Observatory, Avenida Alonso de C\'ordova 3107, Vitacura 7630355, Santiago, Chile}
\affiliation{National Radio Astronomy Observatory, 520 Edgemont Road, Charlottesville, VA 22903-2475, United States of America}

\author{Kazumasa Iwai}
\affiliation{Institute for Space-Earth Environmental Research, Nagoya University, Furo-cho, Chikusa-ku, Nagoya, 464-8601, Japan}

\author{Stephen White}
\affiliation{Space Vehicles Division, Air Force Research Laboratory, Albuquerque, NM, USA}

\begin{abstract}
Polar brightening of the Sun at radio frequencies has been studied for almost fifty years and yet a disagreement  persists between solar atmospheric models and observations. Some observations reported brightening values much smaller than the expected values obtained from the models, with discrepancies being particularly large at millimeter wavelengths.
New clues to calibrate the atmospheric models can be obtained with the advent of the Atacama Large Millimeter/submillimeter Array (ALMA) radio interferometer. In this work, we analyzed the lower limit of the polar brightening observed at 100 and 230 GHz by ALMA, during its Science Verification period, 2015 December 16-20. We find that the average polar intensity is higher than the disk intensity at 100 and 230 GHz, with larger brightness intensities at the South pole in eight of the nine maps analyzed. 
The observational results were compared with calculations of the millimetric limb brightnening emission for two semi-empirical atmospheric models, FAL-C \citep{FAL1993} and  SSC \citep{Selhorst2005a}. Both models presented larger limb intensities than the averaged observed values. The intensities obtained with the SSC model were closer to the observations, with polar brightenings of 10.5\% and  17.8\% at 100 and 230 GHz, respectively.  This discrepancy may be due to the presence of chromospheric features (like spicules) at regions close to the limb.    
\end{abstract}


\section{Introduction} 

In the solar atmosphere, a positive temperature gradient above the photospheric minimum temperature is expected \citep[e.g.][]{VAL1981,FAL1993}, predicting an increase in brightness temperature from millimetric to centimetric wavelengths.
To verify this prediction, many successful observations have been reported at frequencies between 17 and 860~GHz \citep[see ][and references therein]{Selhorst2003}, with generally consistent results. In addition to the limb brightening, the radio observations at polar regions showed an intrinsic behavior, that was designated as polar brightening. 

The great part of our knowledge about the polar brightening comes from the interferometric maps taken since 1992 at 17~GHz  by the Nobeyama Radioheliograph \citep[NoRH]{Nakajima1994}. Based on these maps, \cite{Shibasaki1998} reported that regions near the solar poles showed brightness temperature values up to 40\% above the quiet Sun values, in contrast with equatorial limb regions having only 10\% increase. This suggested that the presence of the polar brightening is associated with bright patches near the solar poles. 

The 17~GHz polar brightening follows the faculae cycle \citep{Selhorst2003} and polar magnetic field strength  \citep{Selhorst2011,Shibasaki2013}, which is anti-correlated with the sunspot cycle. Although bright patches and faculae occur at the poles, there is no one-to-one correlation between them \citep{Gelfreikh2002}. Nevertheless, they are frequently associated with the location of increased unipolar magnetic regions underlying the coronal holes \citep{Gopal1999,Brajsa2007,Selhorst2010}. This association was reinforced by \cite{Oliveira2016}, who analyzed the polar behavior between 2010 and 2015 and found a good correspondence between the presence of coronal holes and polar brightening at 17~GHz.

The radio polar brightening is presumed to result from the atmospheric structure in this region. Observations at additional wavelengths are valuable since they probe different layers of the atmosphere: in particular, millimeter wavelengths probe deeper into the chromosphere. However, only few observations at millimeter/submillimeter wavelengths were performed and presented inconclusive results, that can be partially caused by the different antenna resolutions and the observational techniques employed. In the millimetric range, \cite{Pohjolainen2000b} analyzed the standard right ascension and declination maps observed by the Mets\"ahovi antenna at 87~GHz (half-power beam width, $HPBW=60''$) and reported that polar regions presented  brightening increase of 0.5--2\%. \cite{Kosugi1986} used radial scans obtained with Nobeyama 45-m telescope, and reported no polar brightening at 98~GHz (HPBW= $17''$), but detected a polar brightening of  3--7\% at 36~GHz (HPBW= $46''$), suggesting that the polar brightenings should be an upper chromospheric feature. 

Approaching the sub-millimeter range, \cite{Horne1981} performed radial scans of the Sun with a single antenna of OVRO at 230~GHz ($HPBW= 28''$)  and reported polar brightenings of $10\%\pm5\%$. 

Previous studies based on single-dish observations used radio telescopes with a large amount of broad sidelobes. Therefore, the limb study requires careful deconvolution of the sidelobe \citep[e.g. ][]{Iwai2015,Iwai2017}.

Solar observations with the {\em Atacama Large Millimeter-submillimeter Array} \citep{Wedemeyer2016} have started within its Cycle 4 of observations. The new observations at 100 and 230~GHz have already provided new insights about the solar atmosphere \citep{White2017,Brajsa2018}, indicating a great potential for improvement in our understanding of the Sun.

Aiming to verify the presence of the polar brightening in the millimetric range, in this work we analysed ALMA single-dish maps observed at 100 and 230~GHz during the Science Verification period in December 2015.   The same data set was used by \cite{Alissandrakis2017} to investigate the center-to-limb variation of the quiet Sun, however, the paper did not investigate the 
issue of the polar brightening, which is the specific focus of our paper.

As a secondary goal we also measured the solar radius using the ALMA solar maps. The photospheric radius has been measured for centuries while the measurements at radio wavelengths started only some decades ago. The solar radius is an important parameter for the calibration of solar atmospheric models and for a better understanding of the changes in the atmospheric structure. These measurements also show the altitude where most of the emission at the observed frequencies are generated.

We measured the polar brightening and estimated the solar radius. The results were compared with the solar atmospheric models proposed by \cite{FAL1993} and \cite{Selhorst2005a} and with previous observations at millimetric wavelengths.

\section{ALMA solar maps}

\begin{figure*}[ht!]
\centering
\includegraphics[width=18cm]{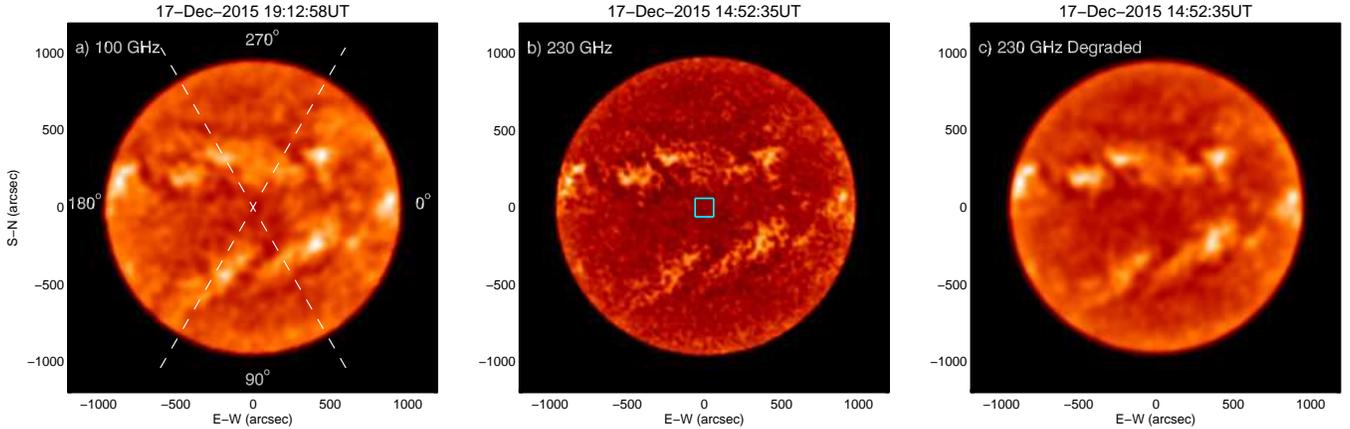} 
\caption{ALMA single-dish maps at 100 (a), 230~GHz (b) and the 230~GHz degraded to the 100~GHz resolution. The dashed lines in the panel (a) delimite the polar regions analyzed in this study, while the box in panel (b) set the region used to calculate the quiet Sun temperature.}
\label{fig:maps}
\end{figure*}

We analysed ALMA fast-scan single-dish maps described by \cite{White2017}. The data were observed during the Science Verification period, 2015 December 16-20, and released in the beginning of 2017\footnote{https://almascience.eso.org/alma-data/science-verification}. The data comprise six maps obtained in Band 3 (84 --116 GHz) and other three in Band 6 (211 -- 275 GHz). {In this work, we took 100 and 230~GHz as the reference frequencies for Bands 3 and 6, respectively}. Since, these maps were obtained by a single dish antenna with 12~m of diameter, they present nominal spatial resolutions of 58'' at 100 GHz and 25'' at 230 GHz \citep{White2017}. Figure~\ref{fig:maps} shows examples of the ALMA single dish maps obtained at 100~GHz (a), 230~GHz (b) and the 230~GHz degraded to the same spatial resolution of the 100~GHz maps.  The brightness temperature values, $T_B$, were scaled as $T_B^4$ to increase the contrast of features on the disk. 

\section{Data analysis and results}

The polar brightening analysis was made in terms of the relative intensity above the quiet Sun brightness temperature ($T_{qS}$). Following the method used by \cite{White2017}, the $T_{qS}$ of each map was assumed to be the mean brightness temperature of a small area at disc center ($120''\times120''$, see Figure~\ref{fig:maps}b), which is not affected by active regions. The obtained values and their respective errors are listed on Table~\ref{table1}. The $T_{qS}$ at Band 3 maps presented a substantial variation between the observations made on the December 16 and 17, which may have affected the polar brightening results. 

The maps obtained at 230~GHz were degraded in order to compare the observations with approximately the same resolution (Figure~\ref{fig:maps}c). The degradation process was performed by the convolution of the Band~6 maps and a bi-dimensional gaussian beam with 54'' (18 pixels), that was determined by the relation $\theta_{100}^2=\theta_{230}^2+\theta_x^2$, assuming that the ALMA observational beams are gaussian.

\begin{figure}[!h]
\centering
\includegraphics[width=9cm]{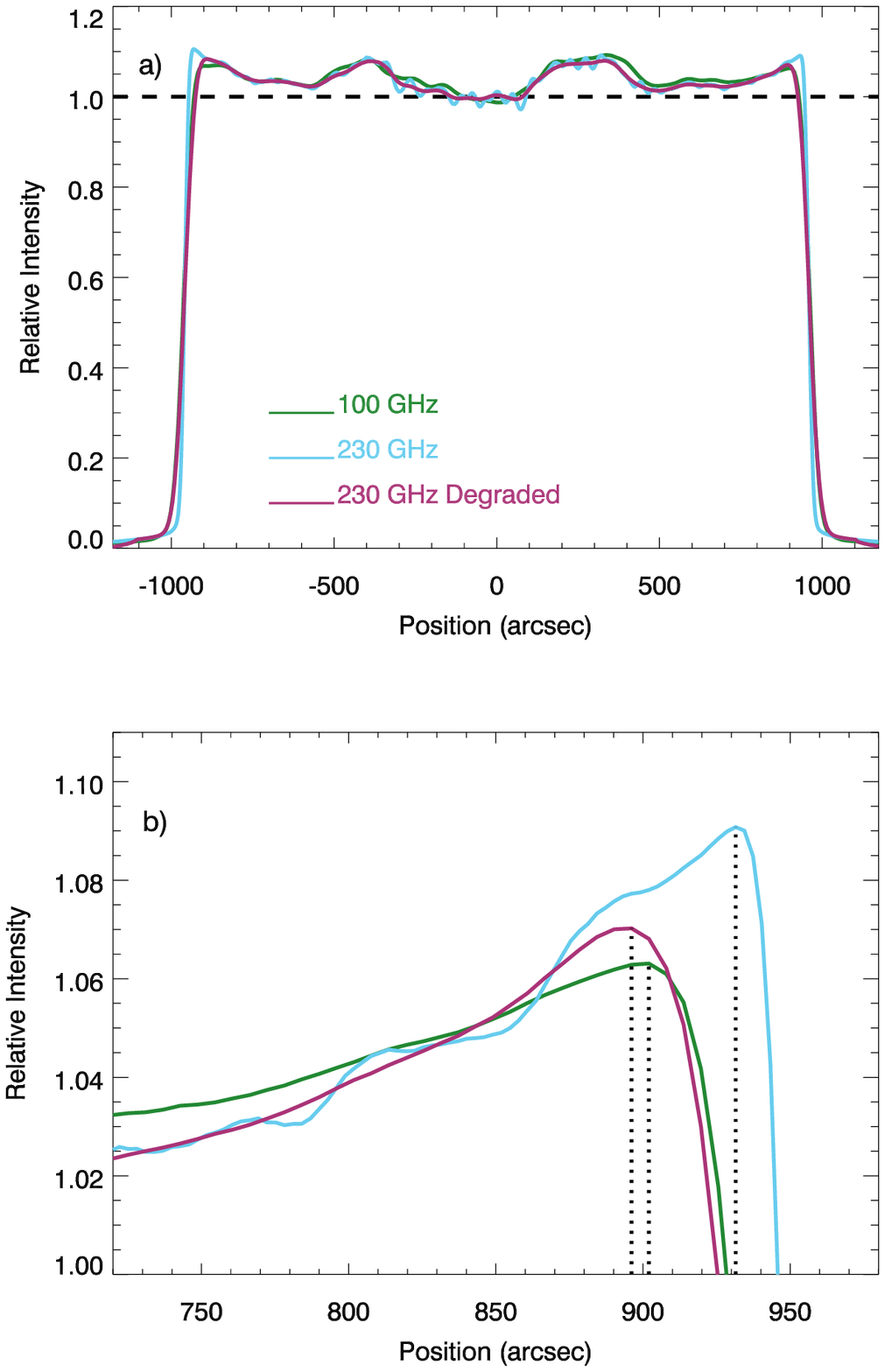} 
\caption{a) Comparison between the averaged polar profiles obtained at 100~GHz (red) and 230~GHz (blue,  and degraded in pink), in which the dashed line represents the quiet Sun intensity. b) Detail of the average North pole profile at both frequencies, as well as the degraded profile. The dotted lines indicate the measured polar brightening position.}
\label{fig:scan}
\end{figure}

Following \cite{Selhorst2003,Selhorst2011}, the polar brightening  was estimated from the average of  61 radial profiles through the disc center, that were taken every 1$^\circ$ between 60 and 120$^\circ$, and between 240 and 300$^\circ$ position angles (corresponding to the South and North poles, respectively). The average profiles for the ALMA maps showed in Figure~\ref{fig:maps} are plotted in Figure~\ref{fig:scan}a, in which the dashed line represents the quiet Sun intensity ($T_{qS}$). 

Figure~\ref{fig:scan}b shows  in detail the average North pole profile (right side in  Figure~\ref{fig:scan}a) at Bands 3 and  6 (original and degraded). The dotted lines in Figure~\ref{fig:scan}b present the measured polar brightening position. 
Due to the large differences in the spatial resolution, the radial position of the peak brightness at 100~GHz ($896''.1$) is $\sim 35''$ closer to Sun center than at 230~GHz ($931''.5$), however, the brightness peak for the 230~GHz degraded profile is almost coincident with the 100~GHz profile. The  difference in antenna beam spatial resolution is also responsible for the  intensity reduction observed at lower resolutions (see Table~\ref{table1}).      

A summary of the results can be seen in Table \ref{table1}, in which the first three columns are the number reference, day and hour of the observation, followed by observed frequency and brightness temperature of the quiet Sun $T_{qS}$. The maximum intensity and the position of the South and North average profiles are presented in the next columns, where the intensity uncertainty values are taken at maxima intensity positions. The last column shows the mean solar radius.

\begin{table*}
\centering
\caption{Average polar brightening and measured solar radius.}
\label{table1}
\begin{tabular}{lcccccccccc}
\hline
\hline
Map \#&Day& Hour (UT) & Freq. (GHz) & $T_{qS}$ (K)& \multicolumn{4}{c}{Polar Brightening (lower limits)} & Solar Radius ($''$)\\
\hline
& & & & & \multicolumn{2}{c}{South}  & \multicolumn{2}{c}{North}  & at $0.5 T_{qS}$\\
\hline
& & & & & Intensity (\%) & Position ($''$) & Intensity (\%) & Position ($''$) & \\
\hline
1& 12/16/2015& 18:27:42 & 100 & $7560\pm 70$ & $4.1 \pm 2.1$& 890.2 & $3.0 \pm 1.8$& 902.0 & $965.3 \pm 3.2$\\
2& 12/16/2015& 19:42:05 & 100 & $7370\pm 70$ & $4.5 \pm 2.1$& 884.3 & $2.9 \pm 1.9$& 902.0 & $965.3 \pm 3.1$\\
3& 12/17/2015& 19:12:58 & 100 & $7100\pm 100$ & $7.0 \pm 2.8$& 854.8 & $6.3 \pm 2.5$& 902.0 & $966.0 \pm 3.3$\\
4& 12/17/2015& 19:20:37 & 100 & $7080\pm 90$ & $6.8 \pm 2.6$& 854.8 & $6.2 \pm 2.4$& 902.0 & $966.1 \pm 3.1$ \\
5& 12/17/2015& 19:28:02 & 100 & $7070\pm 90$ & $7.1 \pm 2.7$& 854.8 & $6.2 \pm 2.4$&896.1 & $966.4 \pm 3.0$ \\
6& 12/17/2015& 19:35:27 & 100 & $7110\pm 90$ & $7.0 \pm 2.5$& 860.7 & $6.2 \pm 2.4$&896.1 & $966.1\pm 3.2$ \\
7& 12/17/2015& 14:52:35 & 230 & $6210\pm110$ & $10.5 \pm 3.4$&931.5 &  $9.1\pm3.4$&931.5 & $961.6 \pm 2.6$ \\
8& 12/18/2015& 20:12:21 & 230 & $6350\pm 210$ & $7.1 \pm 4.7$ &931.5 & $7.9\pm 4.6$& 931.5 &$961.3 \pm 1.7$ \\
9& 12/20/2015& 13:52:39 & 230 & $6270\pm220$ & $10.0 \pm 5.2$ &931.5 & $9.2 \pm 4.8$&931.5 &  $961.9 \pm 2.0$ \\
10& 12/17/2015& 14:52:35 & 230* & $6210\pm50$ & $8.3 \pm 1.8$&890.2 &  $7.0\pm1.4$&896.1 & $964.1 \pm 2.6$ \\
11& 12/18/2015& 20:12:21 & 230* & $6310\pm 60$ & $6.0 \pm 1.9$ &890.2 & $6.6\pm 1.6$& 896.1 &$963.8 \pm 1.4$ \\
12& 12/20/2015& 13:52:39 & 230* & $6270\pm80$ & $7.3 \pm 2.3$ &896.1 & $7.0 \pm 1.8$&896.1 &  $964.4 \pm 2.0$ \\

\hline
\end{tabular} 
\end{table*}  

 As can be verified in Table~\ref{table1}, the 100~GHz maps obtained on December 16th presented  $T_{qS}$ values larger than those observed on December 17th. Nevertheless, since our analysis was based on the relative intensities in each map, we decided to use all maps in the study. The average $T_{qS}$ obtained for the six 100~GHz maps was $7220\pm 200$~K, that is in agreement with the values suggested by \cite{White2017} ($7280\pm250$~K) and \cite{Alissandrakis2017} (7250~K). At 230~GHz, the mean $T_{qS}$ obtained for the three maps was $6280\pm200$~K, which is above the brightness temperature $5900\pm190$~K suggested by \cite{White2017}, however, it is in agreement with the value obtained by \cite{Alissandrakis2017} (6180~K). The degradation of the 230~GHz has a small influence in the $T_{qS}$ mean central temperature, but, as expected, it reduces significantly the standard deviation in which the averaged result was $6260\pm80$~K.  


Except for the 230~GHz map obtained on December 18th, the South pole showed larger intensity values than the North pole, however, these differences are smaller than the standard deviation. The standard deviations of the polar brightening (Table~\ref{table1}) were calculated at the maximum of each average profile {plus the $T_{qS}$ uncertainty}. For the six 100~GHz maps,  the mean polar brightening values are $T_B/T_{qS} = 6.1\%\pm 2.8$ and $5.1\%\pm2.7$, respectively for the South and North poles. Moreover, both maps obtained on December 16th presented $T_{qS}$ larger than those observed on December 17th, which could have  underestimated polar brightening values. Nevertheless, as can be seen in Figure~\ref{fig:comp}a the 100~GHz maps obtained on the same day presented a very similar mean profile.

The average of the three 230~GHz maps yield $T_B/T_{qS} = 9.2\% \pm 4.7$ for the South and $8.7\% \pm 4.4$ for the North pole intensities. Thus, the average polar brightening at 230~GHz  is larger than the one at 100~GHz, probably due to the better spatial resolution. Furthermore, the 230~GHz maximum brightening occurred at the same position, i.e. the distance to the center of the solar disc, at both poles for all maps, which did not occur at 100~GHz. As can be seen in Figure~\ref{fig:comp}a, the South pole observed at 100~GHz on December 17 (maps \#3--6) presented maxima far from the expected limb, that is caused by the presence of bright features in the solar disc, which can not be separated from the limb brightening due to the large beam \citep{Selhorst2017}. 

For the degraded 230~GHz maps the results are compatible with those obtained at 100~GHz inside the standard deviations, in which the mean polar brightening values are  $T_B/T_{qS} = 7.2\% \pm 2.2$ and $6.9\% \pm 1.6$, respectively for the South and North poles.  

The solar radius of each map was also determined to verify the variation caused by the limb brightening and its convolution with the antenna beam size. Following \cite{Costa1999}, the radius was determined at the point in which  the brightness temperature drops to 50\% of the $T_{qS}$ value. The procedure to obtain the radius was similar to the one used in \cite{Selhorst2011}, which  can be resumed as follows: a) determination of $T_{qS}$; b) solar limb position determined by {\tt contour} procedure of IDL at the level equal to half of $T_{qS}$; c) 1440 points determined over the curve resulting from step b; d) a circumference fit adjusted to the points.  The obtained solar radius values presented mean values of $965''.9\pm3''.2$ at 100~GHz and $961''.6\pm2''.1$ at 230~GHz. These measurements are larger than the mean values obtained at the inflection point by \cite{Alissandrakis2017}, but agree within the standard deviations. Moreover, the mean radius at 230~GHz also agrees with the averaged value obtained by \cite{Menezes2017} at 212~GHz ($966''.5 \pm 2.8''$). As expected, the degradation of the 230~GHz maps increases their radii to an averaged value of $964''.1\pm2''.1$.

\section{Atmospheric Models}

In a recent report, \cite{Alissandrakis2017} analyzed the center-to-limb variation of the same set of ALMA maps studied here and suggested that the observations were best fitted by the average quiet Sun model C (FAL-C) proposed by \cite*{FAL1993}. Thus, we calculated the FAL-C brightness temperature variation from the center-to-limb, where the H free-free opacity ($\kappa_\nu{\rm(H)}$), as well as, the $\rm H^-$ free-free opacity ($\kappa_\nu{\rm(H^-)}$) were considered, that can become significant at  {submillimetric wavelengths}. 

\begin{figure}[!h]
\centering
\includegraphics[width=9cm]{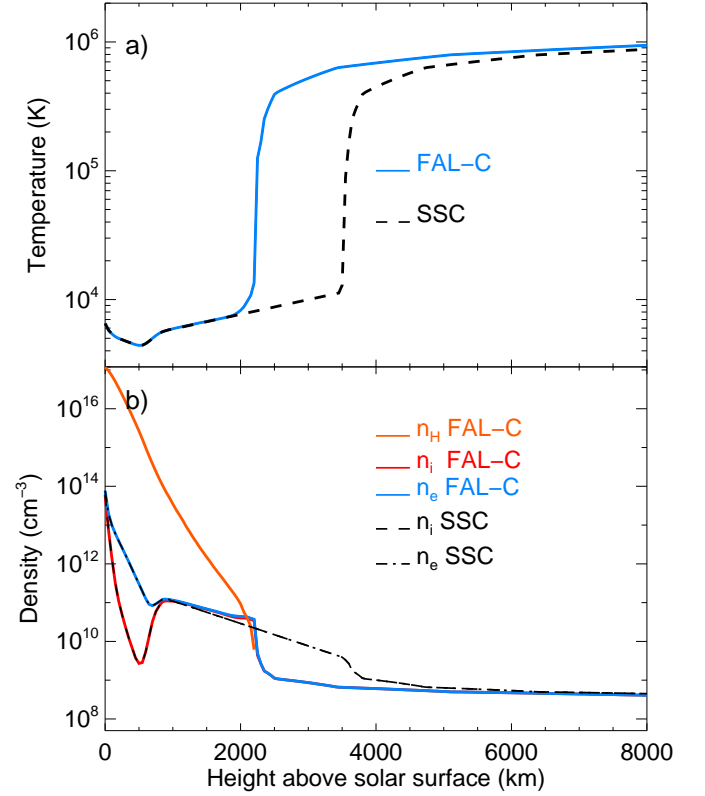} 
\caption{(a) Temperature and (b) densities ($n_e$, $n_i$ and $n_H$) distributions   for the FAL-C and SSC atmospheric models.}
\label{fig:model_1D}
\end{figure}

The observed polar brightenings were also compared with atmospheric model SSC proposed by \cite*{Selhorst2005a}, which follows the FAL-C temperature and densities distributions up to 1000~km above the solar surface, however, for a good fit at radio wavelengths, the authors considered an extended chromosphere, in which the base of the transition region is kept at 3500~km, as in the \cite{Zirin1991} model. This extended chromosphere is proposed to address the mean contribution of the  chromospheric features unresolved by radio observations, such as spicules. The SSC model also considers a coronal plasma up to 40 000~km above the solar surface. The same corona was included in the FAL-C model in order to estimate the coronal contribution to the limb brightening at ALMA Bands 3 and 6 (100 and 230 GHz respectively). Figure~\ref{fig:model_1D} shows the comparison between the temperature and densities distributions proposed by the FAL-C and SSC atmospheric models. It is necessary to take in account that the main purpose of these models is reproduce the brightness temperature of solar disc center and do not considere the special features observed at the polar regions.

The same equations used in \cite{Simoes2017} to calculate $\kappa_\nu{\rm(H)}$ and $\kappa_\nu{\rm(H^-)}$  (in $\rm cm^{-1}$) were adopted here, as follows: 

\begin{equation}
\kappa_\nu{\rm(H)}=3.7\times10^8T^{-1/2}n_en_i\nu^{-3}g_{ff}
\end{equation}

\noindent where the $n_e$ and $n_i$ are the electron and proton densities, respectively, $T$ is the electron temperature and $g_{ff}$ the Gaunt factor,  numerically obtained by \cite{vanHoof2014}.

Following \cite{Kurucz1970}, the $\rm H^-$ free-free opacity is calculated by:

\begin{equation}
\kappa_\nu{\rm(H^-)}=\frac{n_e n_H}{\nu} (A_1+(A_2-A_3/T)/\nu)
\end{equation}

\noindent  where $n_H$ is the neutral hydrogen density, and the numerical coefficients are $A_1=1.3727\times10^{-25}$, $A_2=4.3748\times10^{-10}$, and $A_3=2.5993\times 10^{-7}$. 

The total opacity is given by 

\begin{equation}
\kappa_\nu=[ \kappa_\nu{\rm(H)} +\kappa_\nu{\rm(H^-)}](1-e^{-h\nu/k_bT}),
\end{equation}  

\noindent  with the term $(1-e^{-h\nu/k_bT})$ being the correction for stimulated emission, where $h$ and $k_b$ are the Planck and Boltzmann constants, respectively.

Then, the optical depth is determined by $\tau_\nu=\int \kappa_\nu ds$. The formation height of emission is given by the contribution function (CF, given in $\rm erg s^{-1} cm^{-3}Hz^{-1}sr^{-1}$ throughout this paper), which is defined as

\begin{equation}
{\rm CF}(h)=j_\nu e^{-\tau_\nu}
\end{equation}

\noindent  where $j_\nu=\kappa_\nu B_\nu(T)$ is the emission coefficient, and  $B_\nu(T)$ is the Planck function.

The brightness temperature as a function of the wavelength is calculated as:

\begin{equation}
T_B(\nu)=\int T\kappa_\nu e^{-\tau_\nu} ds.
\end{equation} 

\begin{figure}[!h]
\centering
\includegraphics[width=9cm]{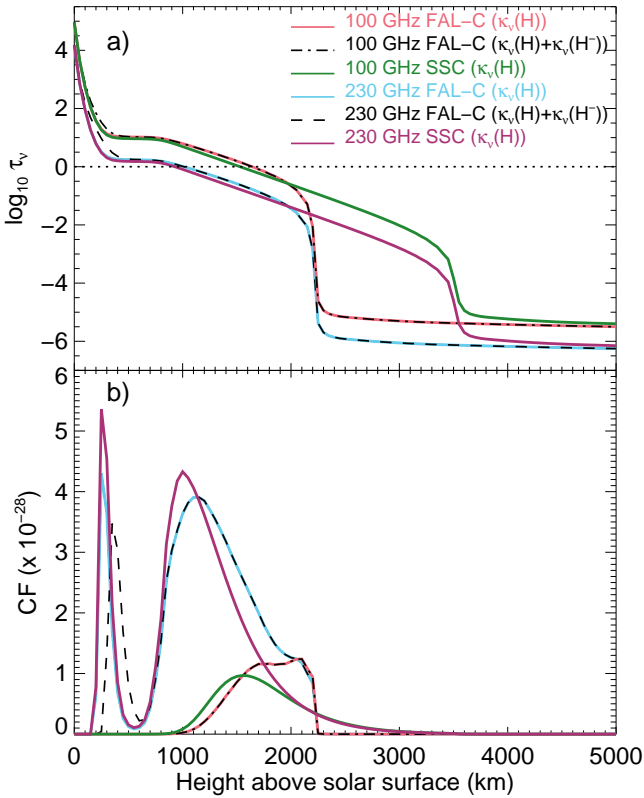} 
\caption{The panel (a) shows the variation of the optical depth ($\tau_\nu$) with the height above the solar surface, in which the horizontal dotted line represents $\tau=1$.  The the variation of the contribution function (CF) with the height above the solar surface is plotted on (b).} 
\label{fig:cf}
\end{figure}

The brightness temperature calculated for the 1D model distributions, plotted in Figure~\ref{fig:model_1D}a and \ref{fig:model_1D}b, are assumed as the $T_{qS}$. If only the free-free opacity, $\kappa_\nu{\rm(H)}$, is considered, the FAL-C model gives $T_{qS}=$7670~K at 100~GHz and $T_{qS}=$6220~K at 230~GHz. The inclusion of the $\rm H^-$ free-free opacity did not change the 100~GHz value, and a small reduction in the value at 230~GHz, namely, 6190~K. For the SSC model ($\kappa_\nu{\rm(H)}$ only),  the results were $T_{qS}=$7180 and 5970~K, at 100 and 230 GHz, respectively. These results are summarized on Table~\ref{table2}. 

The optical depth at 100~GHz is showed in Figure~\ref{fig:cf}a by the curves salmon, green and dot dashed. The $\tau_\nu=1$ layer occurs in the chromosphere around $\sim 1550$~km and $\sim 1650$~km above the solar surface, respectively, in the SSC and FAL-C models. At these heights the opacity is completely dominated by H free-free. {The 100~GHz contribution function CF (Figure~\ref{fig:cf}b) reaches a maximum around 1500~km in the SSC model, and presents relatively constant values between 1650~km and 2100~km in the FAL-C model. Since the $\tau_\nu=1$ layer for the FAL-C model is around 1650~km, it is clear that the emission formed above this height is optically thin.} 

At 230~GHz, inspecting the contribution function CF (Figure~\ref{fig:cf}b), the emission is formed mainly at two different layers: near the temperature minimum region ($h < 400$~km) and in the chromosphere (roughly between $800< h < 1500$~km). The optical depth $\tau_\nu$ at 230~GHz (Figure~\ref{fig:cf}a, red, purple and dashed curves) reaches $\tau_\nu \gg 1$ around photospheric temperature minimum ($h < 400$~km). From the contribution function CF, it is clear that the height of formation of the emission is slightly altered when the H$^-$ opacity is considered. The chromospheric peak of the 230~GHz CF is formed by H free-free radiation, and it is not completely optically thin ($\tau_\nu \sim 1$) nor optically thick: the photospheric contribution is still visible through the chromosphere,  if the observation is made at solar disc center.  

\begin{table*}
\centering
\caption{Atmospheric Models Results}
\label{table2}
\begin{tabular}{lccccccccccc}
\hline
\hline
Model & Freq. (GHz) & $T_{qS}$ (K)& \multicolumn{6}{c} {Limb Brightening} &  \multicolumn{3}{c} {Solar Radius ($''$)}\\
\hline
& & & \multicolumn{3}{c} {Intensity (\%)} &  \multicolumn{3}{c} { Position ($''$)}  & \multicolumn{3}{c} {at $0.5 T_{qS}$} \\
\hline
\multicolumn{3}{c}{HPBW ($''$)}  & 0.14 & 25 & 58 & 0.14 & 25 & 58 & 0.14 & 25 & 58 \\
\hline
FAL-C [$\kappa_\nu{\rm(H)}$] & 100 & 7670 & 79.6 & 23.1 & 15.8 & 962.7 & 939.9 & 909.3 & 962.8 & 966.6 & 970.2\\
FAL-C [$\kappa_\nu{\rm(H)}$ + $\kappa_\nu{\rm(H^-)}$] & 100 & 7670 & 79.6 & 23.1 & 15.8 & 962.7& 939.9 & 909.3 & 962.8 & 966.6 & 970.2\\
SSC [$\kappa_\nu{\rm(H)}$] & 100 & 7180 & 33.7 & 14.6 & 10.5 & 963.2 & 939.5 & 906.6 & 964.3 & 966.7 & 969.1\\
FAL-C [$\kappa_\nu{\rm(H)}$] & 230 & 6220 & 73.5 & 20.2 & 14.7 & 962.5 & 938.5 & 905.7 & 962.8 & 966.0 & 969.1 \\
FAL-C [$\kappa_\nu{\rm(H)}$ + $\kappa_\nu{\rm(H^-)}$] & 230 & 6190 & 74.3 & 20.7 & 15.2 & 962.5 & 938.5 & 905.9 & 962.8 & 966.1& 969.2\\
SSC [$\kappa_\nu{\rm(H)}$] & 230 & 5970 & 37.3 & 17.8 & 13.3 &962.4 & 939.0 & 906.0 & 963.6 & 966.2& 968.9\\
\hline
\end{tabular} 
\end{table*} 

Following \cite{Selhorst2005a}, to calculate the limb brightening, the atmospheric models (Figure~\ref{fig:model_1D}a and \ref{fig:model_1D}b) were expanded as 2D matrices, representing one quadrant of the Sun by taking into account the solar curvature ($R_\Sun=696$~Mm or $959''.64$). Each matrix pixel represents an area of $\rm50\times 50~km^2$. To verify the sharpness of the limb brightening, the brightness temperature was computed every 100~km in the center-to-limb direction,  and these results convolved with a Gaussian function with the same resolutions of the observed maps, i.e., $25''$ and  $58''$.  

The FAL-C high-resolution calculation yields very intense limb brightenings, with a peak of  79.6\% at  100~GHz, for both scenarios, i.e., considering or not the $\rm H^-$ absorption.  Due to their sharp aspect, after the $ 25''$ and  $ 58''$ Gaussian-beam convolution, these  maximum limb brightening values were severely reduced to 23.1 and 15.8\%, respectively. At 230~GHz, the inclusion of the $\rm H^-$ absorption does not change the maximum brightness temperature value ($T_{B_{max}}$), however, due to the small reduction in the $T_{qS}$ caused by its inclusion, the maximum percentages do show a small diference, namely, 73.5 and 74.3\%, however, after the beam convolution,  the intensities were reduced to 20.2 and 20.7\% for a $25''$ beam and 14.7 and 15.2\% for $58''$ beam. In the raw profiles, maximum $T_{B_{max}}$ occurred at the transition region base for both frequencies, that caused an abrupt intensity decrease after the maxima, dropping from the maximum value to $0.5T_{qS}$  in less than 200~km.   

On the other hand, the high-resolution SSC calculations presented smaller maximum temperature values than the FAL-C results. Moreover, the SSC calculations at 230~GHz show larger $T_B/T_{qS}$ limb values (37.3\%) than at 100~GHz (33.7\%), which is the opposite of the high-resolution results for FAL-C. For both Bands, the SSC limb maximum occurred in the middle of the prolonged chromosphere proposed in the model, which resulted in a gradual $T_B$ decrease after the maximum at both Bands. The SSC  {brightness temperature values} dropped from the $T_{B_{max}}$ to $0.5T_{qS}$ in $\sim900$~km. 

Given the more gradual spatial decrease in brightness temperature in the SSC model, after the $T_{B_{max}}$, the Gaussian beam convolution resulted in smaller reduction of $T_B$  {in comparison with} FAL-C simulations. After the convolution with a $25''$ beam, the SSC limb brightening intensities were 14.6 and 17.8\% above the $T_{qS}$, respectively, at 100 and 230~GHz. The convolution with a $58''$ Gaussian-beam resulted in limb intensities of 10.5\% at 100~GHz and 13.3\% at 230~GHz.     

\begin{figure}[!h]
\centering
\includegraphics[width=9cm]{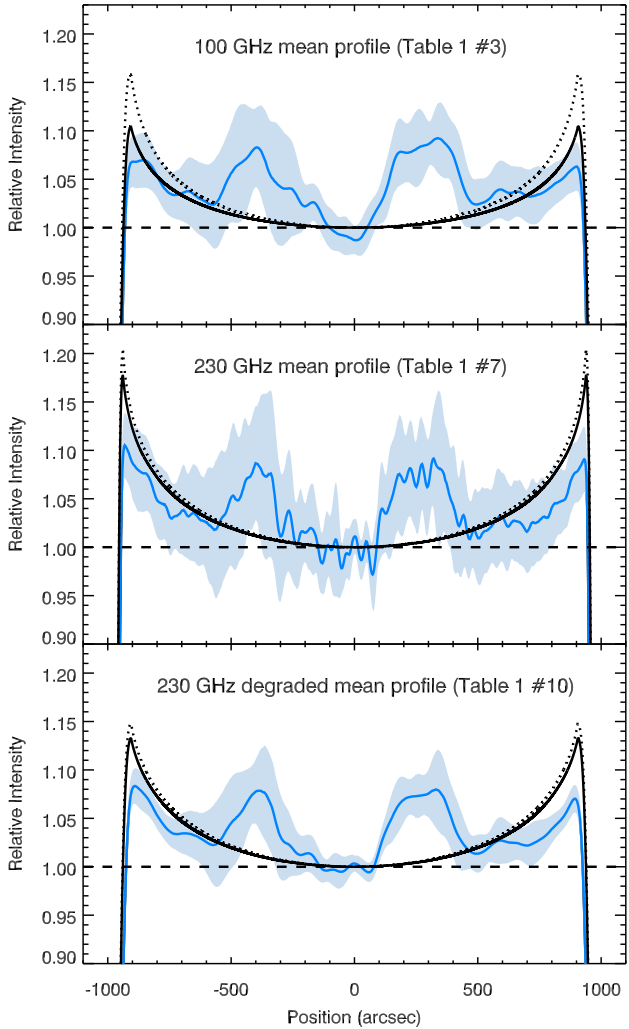} 
\caption{ Comparison between observed brightness temperature mean profiles (blue curves)  and the atmospheric models FAL-C (dotted lines) and SSC (continuous black lines).  The dashed line represents the quiet Sun. The light blue shade around the observational mean profile represents the standard deviation of each point, that includes the $T_{qS}$ uncertainty.}
\label{fig:comp}
\end{figure}

 In Figure~\ref{fig:comp}, we compare the profiles obtained for the models SSC and FAL-C (only $\kappa_\nu(\rm H)$) and the observed $T_B$ mean profiles obtained from the ALMA maps  \#3, \#7  and the degraded \#10 on Table~\ref{table1}. The light blue shadow around the observational mean profile represents the standard deviation of each point, that includes the $T_{qS}$ uncertainty.
Regarding the 100~GHz model,  the SSC and the FAL-C profiles detached from one another toward the limb (Figure~\ref{fig:comp}a), at 230~GHz, the SSC profile mimics  the FAL-C with a smaller $T_B$ (Figure~\ref{fig:comp}b and \ref{fig:comp}c).

 The more intense and larger limb brightening obtained in FAL-C also reflects on the position of the limb brightening maximum and the solar radius at $0.5T_{qS}$. As shown on Table~\ref{table2}, for the raw SSC simulations at 100~GHz the maximum $T_b$ position occurred at $963''.2$, whereas FAL-C presents the maximum $0''.5$ smaller, $962''.7$. Nevertheless, after the Gaussian-beam convolution the positions are reversed, for the $58''$ beam the SSC maximum occurred on $906''.6$, while for the FAL-C, the maximum was located on $909''.3$. The 100~GHz SSC radius increased from $964''.3$ on the raw simulation to $969''.1$ after the convolution with the $58''$ beam, while on the FAL-C, the radius increased from  $962''.8$ to $970''.1$. The same behavior was detected on the 230~GHz radius.

In comparison with the observations (see Figure~\ref{fig:comp}), both models presented larger $T_{B_{max}}$ than the observed mean values at both frequencies. Still, the shape of the simulated center-to-limb profiles generally agrees with the observations when the uncertainties are considered.

\begin{figure}[ht!]
\centering
\includegraphics[width=9cm]{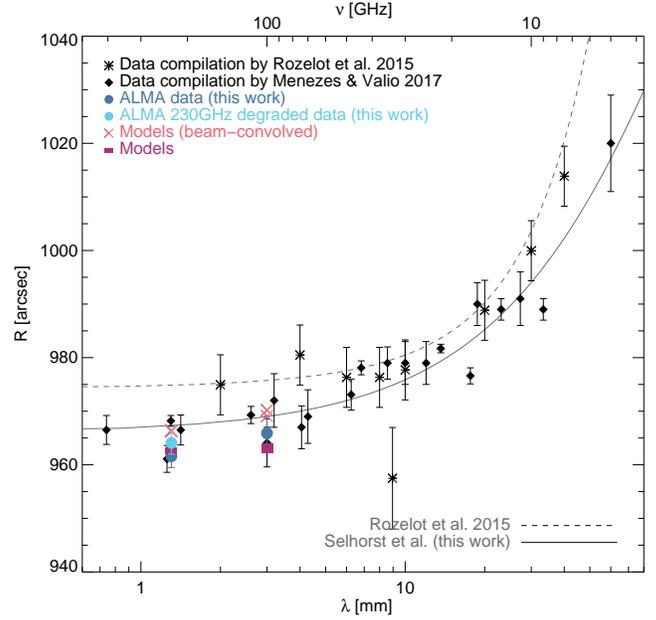} 
\caption{Mean solar radius from measurements at 100 and 230~GHz with other radio measurements collected by \cite{Rozelot2015} and \cite{Menezes2017}.  These historical data were obtained with distinct spatial resolutions and methods. The radii obtained with the models are also shown. 
\label{fig:radius}}
\end{figure}

\section{Discussion and conclusions}

In this work, we report the analyses of the polar brightening observed in the ALMA single dish maps at 100 and 230~GHz. Although the sidelobe level of the ALMA total power antenna should be lower than that of the previous antenna, there should be non-zero sidelobes in a radio telescope beam. Therefore, the derived polar  brightening  should be a lower limit.  The polar brightenings at 100~GHz are $T_B/T_{qS} = 6.1\%\pm 2.8$ and $5.1\%\pm2.7$, respectively for the South and North poles. At 230~GHz, the mean values are $T_B/T_{qS} = 9.2\% \pm 4.7$ for the South and $8.7\% \pm 4.4$ for the North poles. For the 230~GHz maps degraded to the 100~GHz resolution, the polar brightenings reduced to $7.2\%\pm 2.2$ and $6.9\%\pm1.6$, respectively for the South and North poles.  The South pole showed larger brightening values, which may reflect the presence of a small coronal hole, which is known to have small bright structures \citep{Gopal1999,Brajsa2007,Oliveira2016}. Nevertheless, the difference between the poles is smaller than the standard deviation and will need more observations to verify the coronal hole influence on the polar brightening at these frequencies.  


In comparison with the previous results at Band 3 (84--116~GHz), the results at 100~GHz are larger than the $0.5-2\%$ at 87~GHz reported by \cite{Pohjolainen2000b}.  At Band 6 (211 -- 275 GHz), our results are in agreement with those observed by \cite{Horne1981} ($10\%\pm5\%$) at 230~GHz.  


 The averaged solar radii obtained were $965''.9\pm3.2$ at 100~GHz, $961''.6\pm2.1$ for the original 230~GHz maps and $964''.1\pm2.1$ after their degradation. These values agree with those deduced from the inflection point method used by \cite{Alissandrakis2017}, and also agree with the average value obtained by \cite{Menezes2017} at 212~GHz, within standard deviations.  The increase 230~GHz radius after the degradation reflects the great contribution of the intense limb brightening. If it was absent, the radius obtained at $0.5T_{qS}$ should not be dependent of the antenna beam resolution.
 
 In Figure~\ref{fig:radius} we combine the historical data compiled by \cite{Rozelot2015} and \cite{Menezes2017}, also including our own results from ALMA observations.  Above $\approx$50~GHz, our results, along with the recent results compiled by \cite{Menezes2017}, suggest smaller radii at these frequencies than previously suggested by \cite{Rozelot2015}. The continuous curve in Figure~\ref{fig:radius} shows our simple polynomial fit (order 2) to all data shown, while the dashed curve represents a similar curve obtained by \cite{Rozelot2015}. Although these curves may emphasize the height above the photosphere at which the solar radius is determined,  due to the distinct spatial resolution that the radii were obtained, we do not claim that they represent any physical characteristics of the solar radius but are proposed as a simple visualization guide of the typical values of the radii measured at radio frequencies.

The observational results were contrasted with the atmospheric models FAL-C \citep{FAL1993} and SSC \citep{Selhorst2005a}. For the FAL-C model, the $\rm H$ and $\rm H^-$ free-free opacities were considered. The results showed that the $\rm H^-$ absorption does not affect the quiet Sun emission at 100~GHz but changes the emission at 230~GHz in only 30~K. The $T_{qS}$ obtained by FAL-C,  7670~K at 100~GHz,  is higher than the upper limit of the observed values obtained here ($7220\pm200$~K). At 230~GHz the model resulted in 6220~K or 6190~K (with $\rm H^-$), which are in agreement with the observational result $6280\pm 200$~K, but higher than the upper limit of that suggested by \cite{White2017} ($5900\pm190$~K). The $T_{qS}$ overestimation  by the FAL-C at radio frequencies has been previously reported in the centimeter range between 1 and 18~GHz \citep{Bastian1996}.  On the other hand, the SSC presented $T_{qS}=7180$~K at 100~GHz and $T_{qS}=5970$~K at 230~GHz, in agreement with the values suggested by \cite{White2017}  and the observed values at 100~GHz. However, the mean $T_{qS}$ value observed  at 230~GHz is larger than that obtained in the observations.  

All atmospheric models predicted limb brightening intensities larger than the observed ones. However, when the beam sizes were considered, the SSC model yields values closer to the observations, 10.5\% at 100~GHz and  17.8 and 13.3\% at 230~GHz, respectively, for the original and degraded maps. When compared with  observed mean brightness profiles, both models presented are in better agreement with the observations at 230~GHz (Figure~\ref{fig:comp}b). At 100~GHz, however, the FAL-C limb brightening is much broader than the SSC model and the observed profiles.  These discrepancies between the observational and model results can be caused by the different spatial resolutions and sidelobe levels of the telescopes. Since, we considered only the main beam of the telescope and observational result should contain the sidelobe effect, if the sidelobes could be convolved to the atmospheric models, or even removed from the observations, the agreement between the model  and the observational results, will be probably better.

The intense limb brightenings obtained by the models also enlarge the solar radius estimated at $0.5T_{qS}$, making them larger than the observed ones. Besides their larger values, the modelled radii are in agreement with the previous observations showed on Figure~\ref{fig:radius}. The recent data included on Figure~\ref{fig:radius} resulted in a new adjusted curve with smaller values than that one suggested by \cite{Rozelot2015}. The historical results included in Figure~\ref{fig:radius} were measured at different phases of solar cycle, which also may affect the radius at radio frequencies \citep{Selhorst2011,Menezes2017}.  In any case, facing the difficulty to obtain the radio observation data, together with their calibration, the results obtained so far show a rather good agreement with the model and deserves to be described.

Another diference between the observations and models is the position of the maximum limb brightening. At 230~GHz the maximum position is similar for all maps at both poles, located at $931''.5$ from the disc center, which is $\sim 7''.5$ smaller than the modelled position ($\sim 939''$). On the other hand, at 100~GHz, the modelled positions varies from $906''.6$ for the SSC and $909''.3$ for FAL-C, while the observations at the North pole showed peak positions of $896''$ or $902''$ (the difference is only the pixel size), and $\sim 855''$ at the South pole. Such smaller radius measurements are probably due to the presence of bright features close to the limb, that are merged with the usual limb  brightening \citep{Selhorst2017}. Moreover, the limb location may also be affected by the difference of the sidelobes of the telescopes. 

 Apart of the discrepancy with the models, our results are the best we can do with these data but higher resolution data may produce refined 
results for the location and magnitude of any brightening.

 Finally, the observational results listed above are in agreement with those ones reported by \cite{Alissandrakis2017}. However, the SSC model presented limb brightening temperatures closer to the observations, than the values obtained with the FAL-C, which should have the best observational fit between the FAL models \citep{Alissandrakis2017}.    The SSC limb brightening temperature reduction may be due to the prolonged chromosphere, which represents a mean contribution of the unresolved features, like spicules, that guide the chromospheric plasma to coronal heights. This more extended chromosphere contributes to absorption of the sub-mm emission formed at the base of the chromosphere reducing the brightness temperature. Moreover, the presence of these chromospheric features at the regions close to the limb can hardly change the mean temperature profile presented here \citep{Selhorst2005a,Selhorst2005b,Selhorst2017} and will be simulated in a future work. Further investigation can be possible by the modeling of the broad sidelobe \citep[e.g.][]{Iwai2017} and implementation of the sidelobe deconvolution.

\acknowledgments

P.J.A.S. acknowledges support from the University of Glasgow's Lord Kelvin Adam Smith Leadership Fellowship. A.V. and F.M. acknowledge partial funding from Fapesp. R. B. acknowledges partial support by Croatian Science Foundation under the project
6212 ``Solar and Stellar Variability''. This paper makes use of the following ALMA data: ADS/JAO.ALMA\#2011.0.00020.SV. ALMA is a partnership of ESO (representing its member states), NSF (USA) and NINS (Japan), together with NRC (Canada) and NSC and ASIAA (Taiwan), and KASI (Republic of Korea), in co-operation with the Republic of Chile. The Joint ALMA Observatory is operated by ESO, AUI/NRAO and NAOJ. The National Radio Astronomy Observatory is a facility of the National Science Foundation operated under cooperative agreement by Associated Universities, Inc.

\end{document}